\documentclass{PoS}

\title{X/$\gamma$-ray emission from hot accretion flows in AGNs}

\ShortTitle{X-ray and $\gamma$-ray emission from AGNs}

\author{\speaker{A.~Nied\'zwiecki},$^a$ F.-G.~Xie$^{bc}$
        and A.~A.~Zdziarski$^d$\\
\llap{$^a$}Department of Astrophysics, University of \L \'od\'z, Pomorska 149/153, \L \'od\'z, Poland\\
\llap{$^b$}Kavli Institute for Astronomy and Astrophysics, Peking University, Beijing 100871, China\\
\llap{$^c$}Key Laboratory for Research in Galaxies and Cosmology, Shanghai Astronomical Observatory,\\
     Chinese Academy of Sciences, 80 Nandan Road, Shanghai 200030, China\\
\llap{$^d$}Copernicus Astronomical Center, Bartycka 18, 00-716 Warsaw, Poland\\
E-mail: \email{niedzwiecki@uni.lodz.pl}, \email{fgxie@shao.ac.cn}, \email{aaz@camk.edu.pl}}

\abstract{
We present preliminary results of our study of the impact of strong gravity
effects on properties of the high energy radiation produced in accretion 
flows around supermassive black holes. 
We refine a model of the X-ray emission from a hot optically-thin  flow
by combining a fully general-relativistic (GR) hydrodynamical description 
of the flow with a fully GR description of Comptonization. We find that
emission from a flow around a rapidly rotating black hole is 
dominated by radiation produced within the innermost few gravitational radii, the
region where effects of the Kerr metric are strong. The  X-ray
spectrum from such a flow depends on the inclination angle of the line of sight to the symmetry axis,
$\theta_{\rm obs}$, with higher $\theta_{\rm obs}$ characterised by a harder slope and a higher cut-off energy.
For a non-rotating black hole, dependence on $\theta_{\rm obs}$ is insignificant.
 
Under the (reasonable) assumption that the equatorial plane of a rotating supermassive black hole is aligned with the 
surrounding torus, these predicted properties may provide a crucial extension of the unified model of AGNs,
allowing to reconcile the model with systematic trends 
reported in a number of studies of the X-ray spectral properties of AGNs
(indicating that type 2 objects are harder than type 1 and that the relative amount of 
the reflected radiation is larger in the latter). On the other hand, the  model with a rapidly rotating
black hole predicts larger apparent luminosities for objects
observed at higher  $\theta_{\rm obs}$, while an opposite property 
(i.e.\ type 1 objects being more luminous than type 2) was revealed in 
the {\it Integral} data.

We investigate also the hadronic $\gamma$-ray emission from hot flows and
we find much higher (by orders of magnitude) $\gamma$-ray luminosities 
than estimated in previous studies. 
If nearby AGNs contain rapidly rotating black holes and weakly magnetized hot flows,
their $\gamma$-ray emission should be detectable by current $\gamma$-ray detectors.
}

\FullConference{The Extreme sky: Sampling the Universe above 10 keV\\
		October 13-17, 2009\\
		Otranto (Lecce) Italy}

\begin{document}

\section{Introduction}
A large amount of the X-ray data, gathered over the last two decades,
indicate that various kinds of the geometry of accretion flows
occur in AGNs, with an optically thick disc extending down to 
the black hole horizon, truncated 
at several tens of gravitational radii or completely absent in the central
region. We focus here on the last case, characteristic, in particular, 
of radiogalaxies. For these objects,
the most likely solution of accretion flow is that of an optically thin,
two-temperature flow, extensively studied in the context of radiatively
inefficient objects.
Models of the X-ray emission from such flows typically use certain approximations,
in particular, any relativistic effects are neglected, in spite of the (most likely)
origin of this emission in the strongly relativistic region close to the black hole
where most of the gravitational energy is dissipated.
Such simplified models do not allow to study some relevant properties 
of the X-ray spectra, discussed below.

Proton-proton collisions in two-temperature flows lead to substantial $\gamma$-ray emission
through neutral pion production and decay. Recently, observations by {\it AGILE} and {\it Fermi} significantly advanced our exploration of the $\gamma$-ray activity of AGNs.
Motivated by this, we consider the hadronic $\gamma$-ray emission in the previously unexplored 
regime of a weakly magnetized flow around a rapidly rotating black hole (the former property 
is supported by the results of numerical simulations of 
accretion driven by magnetorotational instability and the latter by some evolutionary scenarios
of supermassive black holes).

\section{The models}
\label{model}
We consider a supermassive black hole, characterised by its mass, $M$, and angular momentum, $J$,
surrounded by a geometrically thick accretion flow with an accretion rate
$\dot M$. The following dimensionless parameters are used 
in this paper: $r = R / R_{\rm g}$, $a = J / (c R_{\rm g} M)$, $\dot m = \dot M / \dot M_{\rm edd}$,
where  $R_{\rm g} = GM/c^2$, $\dot M_{\rm edd}= L_{\rm edd}/c^2$ and $L_{\rm edd} \equiv 4\pi GM m_{\rm p} c/\sigma_{\rm T}$.

We solve 
equations of the
dynamical structure of the flow and then we use the computed
profiles of density, temperature and velocity field in modeling the leptonic and hadronic radiative processes. 
Our model is similar to our previous study [1], with three major differences:
(1) our dynamical description of the flow is fully GR here,
while in [1] we use non-relativistic hydrodynamical equations;
(2) the procedure (developed in [1]) for finding the self-consistent structure
of the flow with non-local Compton cooling rate has not been implemented
in the model described here yet;
(3) we neglect here direct viscous heating of electrons. We assume the viscosity parameter $\alpha=0.3$ and the magnetic pressure 
equal 1/10th of the total pressure.

The  Comptonization process is modeled using a Monte Carlo method, see [2]. 
The seed photons are generated from synchrotron and bremsstrahlung emissivities
of the flow; their transfer and energy gains in consecutive scatterings are 
affected by both the special relativistic (SR) and gravitational effects.
In computing the hadronic component, we assume that in the plasma rest frame
the spectrum of $\gamma$-ray photons 
 corresponds to the emission of thermal protons population. We model such a thermal emission 
strictly following [3] and then we compute the transfer of
$\gamma$-ray photons in curved space-time.
Absorption of $\gamma$-rays to pair production is negligible for parameters considered in this paper.

We consider two models without an outflow, i.e.\ with constant $\dot m(r) (=\dot m_0)$, 
with extreme values of the spin parameter, $a=0$ and $a=0.998$,
denoted as model $a_0$ and model $a_1$, respectively. For $a=0.998$ we consider also a model with 
an outflow, with radius-dependent accretion rate, $\dot m (r) = \dot m_0 (r/200)^{0.3}$, 
denoted as model $a_{1,o}$. In all models $\dot m_0=0.1$. We
assume $M = 2 \times 10^8 M_{\odot}$, except for Fig.\ 2c where models with $M = 2 \times 10^8 M_{\odot}$ and $5 \times 10^6 M_{\odot}$ are compared.

 \begin{figure}
 \includegraphics[width=.85\textwidth]{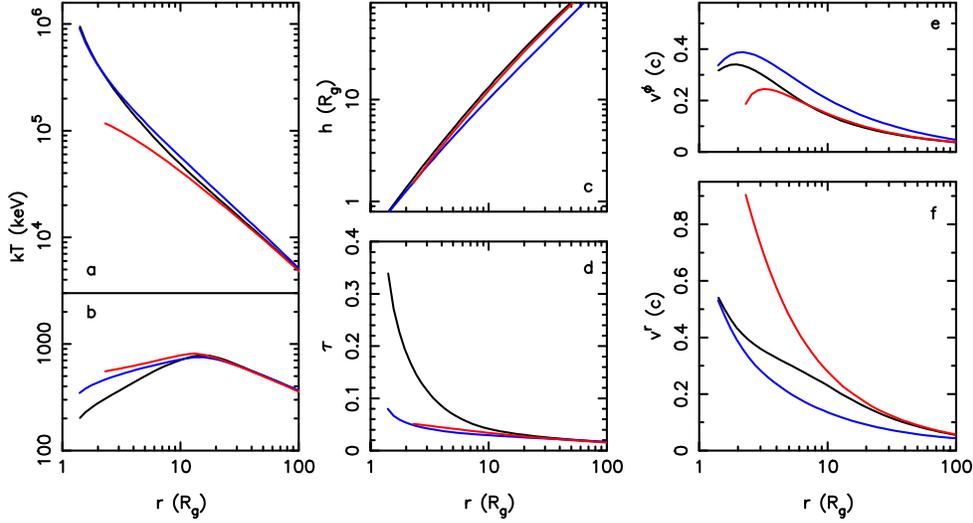}
 \caption{The radial profiles of the proton (a) and electron (b) temperature, 
the half-height (c), the vertical optical depth (d)
and the azimuthal (e) and radial (f) velocities of our hot flow solutions. 
In all panels, the {\it black, blue} and {\it red curves} are for 
model $a_1$, $a_{1,o}$ and  $a_0$, respectively.
}                                                                                                               
\label{fig1}
\end{figure}

\section{Results}

Rotation of the black hole affects 
the structure of the flow in a manner established in previous studies, see Fig.\ 1.
Geometrically thick flows are in general sub-Keplerian,
however, for $a=0$ a rapid radial acceleration occurs 
around $r=10$, while the black hole with $a=0.998$ stabilizes the circular motion of 
the innermost part of the flow and the radial velocity remains rather small down to $r \simeq 2$. 
Then, the continuity equation implies a much
higher value of the optical thickness of the innermost flow around a rapidly rotating 
black hole (note that model $a_0$ 
yields similar $\tau$ as model $a_{1,o}$ although in the latter
most of the matter is lost to the outflow). This, in turn, yields a higher electron temperature 
(required to cool electrons) for $a=0$.
Finally, a more efficient dissipation of gravitational energy occurs in models with $a=0.998$, resulting
in much higher proton temperature at $r<10$ in these models.

These differences in the structure of the central region result in a strong difference 
of the properties of escaping radiation. 
For $a=0.998$ the Comptonized component is much harder 
(by $\Delta \Gamma \simeq 0.2$, $\Gamma$ is the photon spectral index) and the radiative
efficiency (due to electrons only), $\eta$, is an order of magnitude larger  than for $a=0$. Specifically, 
$\eta=0.2\%$ and $0.02\%$ for model $a_1$ and $a_0$,
respectively; for model $a_{1,o}$, $\eta$ is 4 times smaller than in model $a_1$, which 
corresponds to the fraction of the matter lost in the outflow.  

Figs.\ 2(a-c) show the observed spectra from a hot flow, averaged over the ranges of intermediate and small
 $\theta_{\rm obs}$. Figs.\ 2(d-f) give more detailed information on the angular distribution 
of the observed X-ray and $\gamma$-ray fluxes. The dependence on $\theta_{\rm obs}$ results from the combination of the SR
collimation and gravitational focusing 
of radiation. Both effects reduce the flux received by 
face-on observers, more efficiently in high-$a$ models. For $a=0$,
$v^r > v^{\phi}$ and collimation toward the black hole 
dominates over collimation toward the edge-on observers. For high $a$, an effect unique for the Kerr metric,
i.e.\ bending of photon trajectories to the equatorial plane, strongly enhances the flux 
emitted to edge-on directions.

 \begin{figure}
 \includegraphics[width=.75\textwidth]{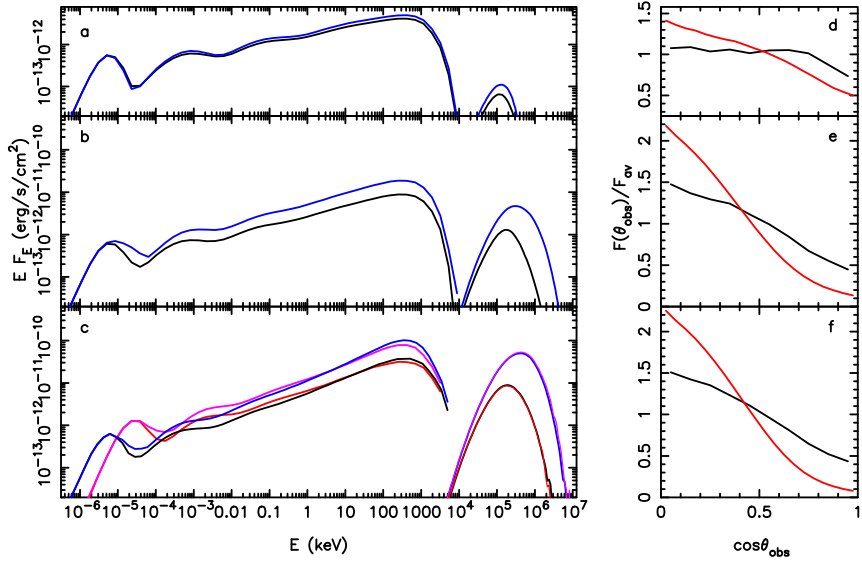}
 \caption{{\it Left panels:} spectra of leptonic (between $10^{-3}$ eV and several MeV) and 
hadronic (between 10 MeV and several GeV) emission from a hot accretion flow.
In all panels, the black and the blue curves
show spectra averaged over all viewing angles with $\cos \theta_{\rm obs} \in [0.7,1]$
and [0.3,0.6], respectively. The assumed distance is $D=10$ Mpc. 
{\it Right panels:} angular distribution of the flux between 2 and 10 keV 
(the black curves) and the total flux in the $\gamma$-ray (hadronic) component (the red curves),
shown as the ratio of the flux observed from a given direction to the observed flux averaged 
over all viewing angles. 
{\it Panels (a,d):} model $a_0$;
{\it panels (b,e):} model $a_{1,o}$;
{\it panels (c,f):} model $a_1$.
Panel (c) shows spectra for both $M=2 \times 10^8 M_{\odot}$ (blue and black) and 
$M=5 \times 10^6 M_{\odot}$ (red and magenta, multiplied by a factor of 40).
}      
\label{fig2}
\end{figure}

The anisotropy is more pronounced for 
the hadronic component, production of which is strongly centrally concentrated and which 
is thus subject to stronger relativistic effects.
For the Comptonized 
radiation, the anisotropy is slightly reduced due to mixing of radiation produced in the innermost part
with that from more distant regions. 

The Comptonized component in model $a_1$ is dominated by radiation produced 
at $r \leq 2$. Radiation from that region has a strong intrinsic anisotropy (see [2]) and
its properties are partially reflected in the shape of the total spectrum,
being slightly harder and having a higher cut-off energy  at larger
$\theta_{\rm obs}$. The effect is rather moderate, with the slopes of the edge-on and the 
face-on spectra differing by   
$\Delta \Gamma = 0.1$ and their cut-off energies differing by a factor of 2. 
For the average face-on and intermediate spectra (shown in Fig.\ 2c; these are appropriate 
for comparison with the stacked spectra of type 1 and 2 objects presented in studies
discussed in Section 4) the difference is even smaller,  e.g.\ $\Delta \Gamma \approx 0.05$.
On the other hand, global cooling effects should reduce the electron temperature at larger $r$ (see [1]),
leading to weaker contribution of the isotropised radiation at high energies; then, a more pronounced dependence of
$\Gamma$ on $\theta_{\rm obs}$  may be expected when these effects are taken into account.

For $a=0.998$, the $\gamma$-ray component contains a  substantial part of the bolometric luminosity
of the flow. E.g., in model $a_1$ the total luminosity of the $\gamma$-ray component is
only by a factor of 3 lower than the luminosity of Comptonized synchrotron and bremsstrahlung
emission. The hadronic production of $\gamma$-rays in hot flows was studied in [4] and [5]
and much smaller $\gamma$-ray luminosities were found.
This was caused by much smaller proton temperatures in their models,
which resulted from the neglect of the black hole rotation in [4]
and the assumed equipartition between magnetic and gas pressures
in [5]. Actually, scaling the results of [4] and [5] to the value of parameters of our models, 
gives the levels of $\gamma$-ray luminosity several times higher than those
in Fig.\ 2, most likely due to the neglect of any relativistic effects affecting the $\gamma$-ray component in both [4] and [5].

\section{Discussion}

Studying effects of strong gravity near black holes is one of the key goals
of high-energy astronomy. So far, direct investigation of such effects
focused on distortions of discrete spectral features emitted from optically thick
discs. We point out that relativistic models of radiative processes in optically-thin flows predict some
observationally testable effects, directly related to the properties
of the space-time of a rapidly rotating black hole. In our discussion of points (i) and (ii)
below, we assume that the midplane of a torus embedding the central region lies 
in the equatorial plane of a central black hole, which is the most natural configuration 
for a rapidly rotating black hole 
 as the spin of the hole is aligned
with the angular momentum of the outer disc on rather short time scale, e.g.\ [6].

{\it (i) Intrinsic anisotropy.} The predicted anisotropy of the X-ray emission  may have crucial implications for 
the unification scheme of AGNs, where various classes of AGNs are supposed to have the same
central engine and the observed differences are attributed to orientation effects.
In the original formulation of this scenario, type 1 and type 2 objects produce 
the X-ray radiation with the same intrinsic spectrum, but the latter are seen at  
larger inclination angles and their emission is partially
absorbed by a torus surrounding the central region. Then, a number of recent reports
that type 2 objects appear to be, on average, intrinsically flatter than type 1, seemed to contradict
the unification model.
We point out that such a (reported) property should indeed characterize the X-ray radiation from AGNs,
if most of them contain rapidly rotating black holes. 
According to our preliminary results (for 
a specific accretion scenario, specific parameters and neglecting global cooling effects),
the GR Comptonization model cannot explain 
the large difference of $\Delta \Gamma \approx 0.4$ between
of the average spectral slopes of Seyfert 1 and Seyfert 2 galaxies, 
derived from the {\it CGRO}/OSSE ([7]) and {\it Swift} ([8]) data. Our predicted $\Delta \Gamma \approx 0.05$ 
approximately agrees  with the average slopes of type 1 and 2 objects found in the {\it Integral} ($\Gamma=1.96$ and 1.91; [9]) and {\it BeppoSAX} ($\Gamma=1.89$ and 1.80; [10])
data with more complex models, involving the reflection component; however, the 
differences of these average spectral slopes are only marginally significant, which may indicate that such a magnitude 
of $\Delta \Gamma$ is too small to be reliably verified with current detectors.
Both [9] and [10] find the average cut-off energies higher in type 2 objects than
in type 1, consistent with our results. On the other hand, the {\it Integral} observations indicate
that the average luminosity is by over a factor of 2 higher for type 1 objects  than for type 2 ([9]),
while exactly an opposite property is predicted in our model. Then, our results are inconsistent
with the scenario in which objects from the observed sample, classified as type 1 or type 2, are characterised by 
the same ranges of parameters (except for viewing angles), most importantly, the same ranges of accretion rates.   

{\it (ii) Relative strength of reflection.}  Another (qualitative) prediction of our model, related
to the intrinsic anisotropy of the X-ray emission,
 is that the amplitude, $\Omega/2 \pi$, of radiation reflected from an optically thick
material should be higher in objects observed at smaller $\theta_{\rm obs}$, for which the ratio
of the directly observed flux to the flux of radiation illuminating the reflecting material (located in the equatorial
direction) would be smaller.
 The {\it Swift} and {\it BeppoSAX} data indeed reveal such a property ([8],[10]), e.g.\ the latter 
give $\Omega/2 \pi=1.23$ for type 1 and $\Omega/2 \pi=0.87$ for type 2 objects.
The {\it Integral} data 
show approximately the same   $\Omega/2 \pi$ for both types of objects ([9]), 
however, a smaller viewing angle was assumed in [9] for the reflection model of type 1 
than for type 2 objects, which actually implies that the reflected component is stronger 
in the former.
  
{\it (iii) $\gamma$-ray emission.} A strong emission of $\gamma$-rays,
with the luminosity comparable to the X-ray luminosity,
may be generated in hot flows around rotating black holes.
The predicted $\gamma$-ray fluxes 
should be easily detectable by {\it Fermi} from nearby (up to a few tens of Mpc) AGNs with parameters 
considered above, specifically, luminosities a few orders of magnitude below
$L_{\rm edd}$ and $M > 10^8 M_{\odot}$ (NGC 1365 or Cen A are interesting examples).
A lack of such a detection would imply that an AGN either contains a slowly rotating black hole 
or its accretion flow is strongly magnetized (both properties significantly
reduce the proton temperature).

\acknowledgments{
AN and AAZ were supported in part by the Polish
MNiSW grants N20301132/1518, 362/1/N-INTEGRAL/2008/09/0 and NN203065933. 
FGX was supported in part by Chinese NSFC (grants 10973003 
and 10843007) and NBRPC (grants 2009CB824800 and 2009CB24901).}

\end{document}